%% file: cfidd_arxiv.tex
\documentclass[onecolumn]{sbrt}
\usepackage[english]{babel}
\usepackage[utf8]{inputenc}

\usepackage[english]{babel}
\usepackage[utf8]{inputenc}
\usepackage{graphicx}
\usepackage{float}
\usepackage{amsmath}
\usepackage{amssymb,color}
\usepackage{algorithm}
\usepackage[noend]{algpseudocode}
    \usepackage{algorithm}
\usepackage{tikz}
\usepackage{pgfplots}
\usepackage{graphicx}
\pgfplotsset{compat=newest}
\usetikzlibrary{shapes,arrows,fit,positioning,calc}
\usetikzlibrary{plotmarks}
\usetikzlibrary{decorations.pathreplacing}
\usetikzlibrary{patterns}
\usetikzlibrary{chains,arrows,shapes,spy}

\usetikzlibrary{automata}
\usepackage{algpseudocode}
\DeclareMathOperator{\tr}{tr}

\DeclareMathOperator{\dBs}{dBs}
\def\BibTeX{{\rm B\kern-.05em{\sc i\kern-.025em b}\kern-.08em
    T\kern-.1667em\lower.7ex\hbox{E}\kern-.125emX}}

\begin{document}

\title{Iterative Detection and Decoding for Cell-Free Massive Multiuser MIMO with LDPC Codes}

\author{Tonny Ssettumba, Roberto B. Di Renna, Lukas T. N. Landau and Rodrigo C. de Lamare
\thanks{Tonny Ssettumba, Roberto B. Di Renna, Lukas T. N. Landau and Rodrigo C. de Lamare, Center for Telecommunications Studies (CETUC), Pontifical Catholic University of Rio de Janeiro (PUC-Rio), E-mail: tssettumba@aluno.puc-rio.br, \{roberto.brauer, lukas.landau, delamare\}@cetuc.puc-rio.br }%
}

\maketitle



\begin{abstract}
This paper proposes an iterative detection and decoding (IDD) scheme
for a cell free massive multiple input multiple output (CF-mMIMO)
system. Users send coded data to the access points (APs), which is
jointly detected at central processing unit (CPU). The symbols are
exchanged iteratively in the form of log likelihood ratios (LLRs)
between the detector and the low-density parity check codes (LPDC)
decoder, increasing the coded system's performance. We propose a
list-based  multi-feedback diversity with successive interference
cancellation (MF-SIC) to improve the performance of the CF-mMIMO.
Furthermore, the proposed detector is compared with the parallel
interference cancellation (PIC) and  MF-PIC schemes. Finally, the
bit error rate (BER) performance of CF-mMIMO is compared with the
co-located mMIMO (Col-mMIMO).
\end{abstract}

\section{Introduction}

Massive multiple-input multiple-output (mMIMO) is a multi-user
communications solution that involves a large number of antennas to
provide service to multiple users in centralized \cite{mmimo,wence}
and distributed \cite{rmmse} settings. The large antenna array
yields high throughput and also improves the propagation conditions
because of the channel hardening property \cite{r1,spa,r3}. mMIMO
leverages on the assumption that users have a single-antenna whereby
there are significantly more antennas at the  Base Station (BS) than
the number of served users \cite{r3}. The signals transmitted by the
users to the receiver overlap, resulting in multi-user interference
at the receiver. These interfering signals cannot be easily
demodulated at the receiver, which call for techniques that can
separate such signals \cite{r4}. The major aim is to reduce the
Euclidean distance between the transmitted signal and the estimate
of the received signal \cite{r5}. Several works have studied optimal
detection techniques to improve the performance of mMIMO.  However,
the complexity of such schemes  increases with the modulation order
and the number of antennas \cite{r5}. Furthermore, sub-optimal
detectors that use iterative detection and decoding (IDD) that
utilise  non-linear techniques such as minimum mean square error
with successive interference cancellation (MMSE-SIC) and parallel
interference cancellation (PIC) have been studied in different works
\cite{r1,r2,r5,r6}. These schemes have been found to achieve close
to optimal  bit error rate (BER) performance.

The key aspect in IDD based strategies is the exchange of soft information between the soft detector and the decoder in terms of likelihood ratios (LLRs).  After some iterations, the decoder sends the interleaved posterior probabilities (extrinsic) information to the soft detector in form of feedback \cite{r6,r7}. The use of codes that use message passing such as low-density parity check codes (LPDC) and turbo codes has been studied in several works  \cite{r8}.

Prior works on IDD that employ channel codes that use message
passing such LDPC and turbo  codes  include the work in \cite{r1,
r2,r3,r4,r5,r6,r7}. Such code designs are less complex which
simplifies communication system. The use of list-based detection
approaches such as:   Multiple-feedback (MF) with SIC (MF-SIC) and
multiple-branch-MF processing with SIC (MB-MF-SIC) detection schemes
have been applied in MIMO architectures to lower the BER
\cite{r5,r6}. Such schemes achieve close to optimal performance and
also reduce the brief error propagation that is prevalent when using
SIC based detection. In \cite{r7}, the uplink of a CF-mMIMO network
has been studied. The access points (APs) are assumed to locally
implement soft MIMO detection and then share the resulting bit LLRs
on the front-haul link without exchanges between the detector and
the decoder. The CPU was used to decode the data while the
non-linear processing at the APs consisted of the approximate
computation of the posterior density for each received data bit.
Moreover, the detection was performed via Partial Marginalization.

In this work, we present an IDD scheme for CF-mMIMO systems, which unlike the work in \cite{r7}, employs message passing. In particular, we propose list-based  MF-SIC detectors based on soft interference cancellation for a centralized CF-mMIMO network. To the best of the authors' knowledge, no such detector has been presented in the previous works for the CF-mMIMO architecture. Moreover, the use of message passing strategies can significantly reduce the BER.  Therefore, the main contributions of this paper are summarized as follows. First, a list-based soft MF-SIC detector is proposed for the CF-mMIMO architecture. This proposed approach gives lower BER values at the same computation complexity as the traditional SIC scheme.  Secondly, the proposed detector is compared with other detectors such as the linear MMSE, SIC, PIC and MF-PIC. Thirdly, the impact of increasing the IDD iterations   is examined. Finally, the CF-mMIMO architecture is compared with the co-located mMIMO (Col-mMIMO) system in terms of the BER performance. The CF-mMIMO significantly achieves lower BER values than the Col-mMIMO.

The rest of this paper is organized as follows:
Section \ref{sys} presents the system model and the statistical analysis. The proposed MF-SIC and MF-PIC detectors are presented in \ref{MFSIC}. Section \ref{IDD} discusses the IDD scheme.  Simulation results and discussions are presented in \ref{Num_Dis}.  Finally, concluding remarks are given  in section \ref{CO_FD}.

\textbf{Symbol notations}: We use lower/upper bold case symbols to represent vectors/matrices, respectively. The Hermitian transpose operator is denoted by $(\cdot)^{H}$.

\section{Proposed System Model} \label{sys}

\begin{figure}[htbp]
\centering
\includegraphics[width=8.5cm]{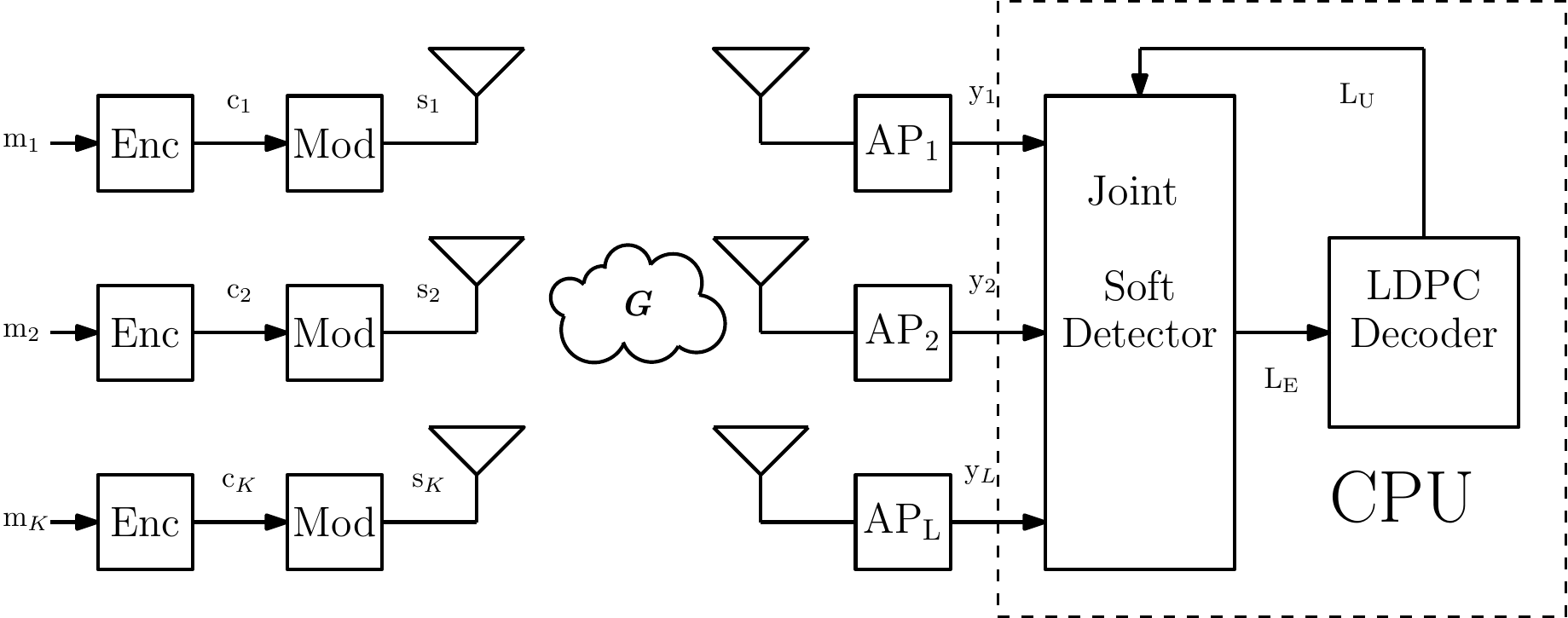}
\caption{Block diagram of a communication system with an IDD scheme.}\label{fig1}
\end{figure}

The  proposed low complexity IDD scheme  for CF-mMIMO systems is
shown in Fig.~\ref{fig1}. Particularly, an LDPC-coded CF-mMIMO
system comprising of $L$ APs, $K$ single antenna user equipments
(UEs), a joint detector at the CPU and an LDPC decoder is
considered.

The data are first  encoded (Enc) by an  LDPC encoder having a code
rate ${R}$. This encoded sequence is then modulated (Mod) to complex
symbols with a complex  constellation of  $2^{M_{c}}$ possible
signal points and average energy $E_{s}$. The coded data is then
transmitted by $K$ UEs through the channel $\mathbf{G}$ to the APs.

We assume a centralized user-centric CF-mMIMO   scenario, where the CPU does soft proceesing and joint detection on the received signal vectors from the APs. Then the CPU sends these soft outputs $L_{E}$ in the form of LLRs to the LDPC decoder. The decoder adopts an iterative strategy by  sending extrinsic information $L_{U}$ to the  CPU which improves the performance of the entire network. Additionally, the  performance of the proposed detector  is examined for the case with no iterations and the case with iterations.
The channel coefficients between the $l$-th AP and the $k$-th UE are given by \cite{r9} \begin{align}
     g_{k,l}=\sqrt{\beta_{k,l}}h_{k,l},
 \end{align}
 where $\beta_{k,l}$ is the large-scale (LS) fading coefficients as a result of path loss  (PL) and shadowing. The small scale fading coefficients are given by $h_{k,l}$, that are independent and identically distributed (i.i.d.) Gaussian random variables with  variance $\mathbb{E}\{ h^{*}_{k,l}h_{k,l}\}=1$.

The LS fading coefficient are assumed to be deterministic and can be obtained using the three-slope PL model \cite{r9}. More precisely, the PL exponent is $3.5$ if the distance $d_{kl}$ between the k-th UE and l-th AP is greater than $d_1$, equals $2$ if $d_{1}\geq d_{kl}>d_{0}$, and equals $0$ if $d_{kl}\leq d_{0}$, for some $d_{0}$ and $d_{1}$. For $d_{kl}>d_{1}$, the Hata-COST231 propagation model is applied.
 The PL  $PL_{kl}$ in ${\dBs}$ between the $k$-{th} UE  and $l$-th AP can be given such as
 \begin{equation} \text {PL}_{kl} \!=\! \begin{cases} -\Lambda - 35\log  (d_{kl}), ~ d_{kl}>d_{1}\\ -\Lambda - 15\log  (d_{1}) - 20\log  (d_{kl}),\qquad   ~ d_{0}< \!d_{kl}\leq d_{1}\\ -\Lambda - 15\log  (d_{1}) - 20\log  (d_{0}), ~ \text {if} ~ d_{kl} \leq d_{0}\\ \end{cases}. \label{PLdB}
 \end{equation}
The parameter $\Lambda$ is given by
\begin{align} \Lambda\triangleq&46.3+33.9\log _{10}(f)-13.82\log _{10}(h_{\text {AP}}) \\\notag&-\, (1.1\log _{10}(f)-0.7)h_{\text {u}}+(1.56\log _{10}(f)-0.8), \\&\notag \end{align}
where $f$ is the carrier frequency (in MHz), $h_{u}$ and $h_{AP}$ are the antenna heights of the UE and AP, respectively. The LS coefficient $\beta_{kl}$ models the PL and shadow fading that is given by
\begin{align}
    \beta_{lk}=PL_{kl}\times10^{\sigma_{sh}\zeta_{lk}}.
\end{align}
Where $10^{\sigma_{sh}\zeta_{l k}}$ denotes the shadowing with standard deviation $\sigma_{sh}$,  and  $\zeta_{l k}\sim{N}(0,1)$.
 The received signal $\mathrm{\mathbf{y}}$ at the joint soft detector is given by
 \begin{align}
 \mathbf{y}=\mathbf{G}\mathbf{s} +\mathbf{n},
 \end{align}
 where $\mathrm{\mathbf{G}}$ $\in C^{L\times K}$ is the channel matrix comprising of both small scale and LS fading coefficients. $\mathbf{s}=[s_{1}, s_{2},..,s_{k-1}, s_k,  s_{k+1},...,s_{K}]$, $\mathbf{n}$ is the additive white Gaussian noise sample (AWGN) with zero mean and unit variance.

\subsection{MMSE soft cancellation detectors}

For simplicity of analysis, we consider sub-optimal detectors which
consists of  PIC/SIC followed by an MMSE filter. The detector first
forms soft estimates of the transmitted symbols by computing the
symbol mean $\bar{s}_{j}$ based on the available a-priori
information from the decoder \cite{r3}
\begin{align}
  \bar{s}_{j}=\sum_{s\in {A}}s P(s_{j}=s),
\end{align}
where ${A}$ is the complex constellation set. By assuming
statistical independence of bits within the same symbol as in
\cite{r3}, the a-priori probabilities are calculated from the
extrinsic LLRs provided by the LDPC decoder as
\begin{align}
\label{aprior_prob}
P(s_{j}=s)=\prod_{l=1}^{M_{c}}\lbrack
1+\exp(-s^{b_{l}}L_{c}(b_{(j-1)M_{c}+l}))\rbrack^{-1},
\end{align}
where $s^{b_{l}}\in (+1,-1)$ denotes the value of the $l$-th bit of
symbol $s$,  $L_{c}(b_{i})$  denotes the extrinsic LLR of the $i$-th
bit computed by the LDPC decoder in the previous iteration. We
define $L_{c}(b_{i}) = 0$  at the first iteration since the only
available belief is from the channel. For the $k$-th  user, the soft
interference from the other $K-1$ users is canceled according to PIC
to obtain
\begin{align}\label{equ8}
\textbf{y}_{k}&=s_{k}\textbf{g}_{k}+\sum_{j=1,j\neq
k}^{K}(s_{j}-\bar{s}_{j})\textbf{g}_{j}+\textbf{n}.
\end{align}
For SIC, the soft interference from the other $K-1$ users is
canceled to obtain
\begin{align}
\textbf{y}_{k}&=\textbf{y}-\sum_{j=1}^{K-1}\bar{s}_{j}\textbf{g}_{j}.
\end{align}
Using \eqref{equ8}, a symbol estimate $\hat{{s}}_{k}$ of the
transmitted symbol on the $k$-th UE is obtained by applying a linear
filter $\textbf{w}_{k}$ to $\textbf{y}_k$ such as
\begin{align}
\hat{s}_{k}&=\mathbf{w}_{k}^{H}\mathbf{y}_{k}\\&\notag=(\mathbf{w}_{k}^{H}\mathbf{g}_{k})s_{k}+\sum_{j=1,\medspace
j\neq
k}^{K}(\mathbf{w}_{k}^{H}\mathbf{g}_{j})(s_{j}-{\bar{s}}_{j})+\mathbf{w}_{k}^{H}\mathbf{n},
\end{align}
where $\mathbf{w}_{k}$ is chosen  to minimize   the mean square
error (MSE) between the transmitter symbol $s_{k}$ and the filter
output $\hat{s}_{k}$ and depends on the variance of the symbols used
in the cancellation step \cite{jidf,rsrbd,rsthp,rapa}. Due to paper
size limitation, the estimated symbol while using the SIC  can be
obtained using a similar approach applied for the PIC.
    In \cite{r1,r3} it is  shown that the corresponding linear  filter is given by
    \begin{align}\label{equ10}
    \mathbf{w}_{k}=\biggl(\frac{\sigma^{2}_{n}}{E_{s}}\mathbf{I}+\mathbf{G}\mathbf{\Delta}_{k}\mathbf{G}^{H}\biggr)^{-1}\mathbf{g}_{k},
    \end{align}
    with
    \begin{align}\label{delta_eq}
    \mathbf{\Delta}_{k}=\mathsf{diag}\bigg[\frac{\sigma_{s_{1}}^{2}}{E_{s}},...,\frac{\sigma_{s_{k-1}}^{2}}{E_{s}}, 1 , \frac{\sigma_{s_{k+1}}^{2}}{E_{s}},...,\frac{\sigma_{s_{K}}^{2}}{E_{s}}\bigg],
    \end{align}
    \noindent where ${\sigma_{s_{i}}^{2}}$ is the variance of the $i$-th user symbol computed as
    \begin{align}
    \sigma_{s_{i}}^{2} =\sum_{s\in \mathcal{A}}^{}|s-\bar{s}_{i}|^{2}P(s_{i}=s).
    \end{align}

\section{Proposed Multi-Feedback detection-SIC}\label{MFSIC}

In this section, we describe the operation of the proposed
list-based detection scheme
\subsection{MF-SIC Design}
\begin{figure}[htbp]
\centering
\includegraphics[width=7cm]{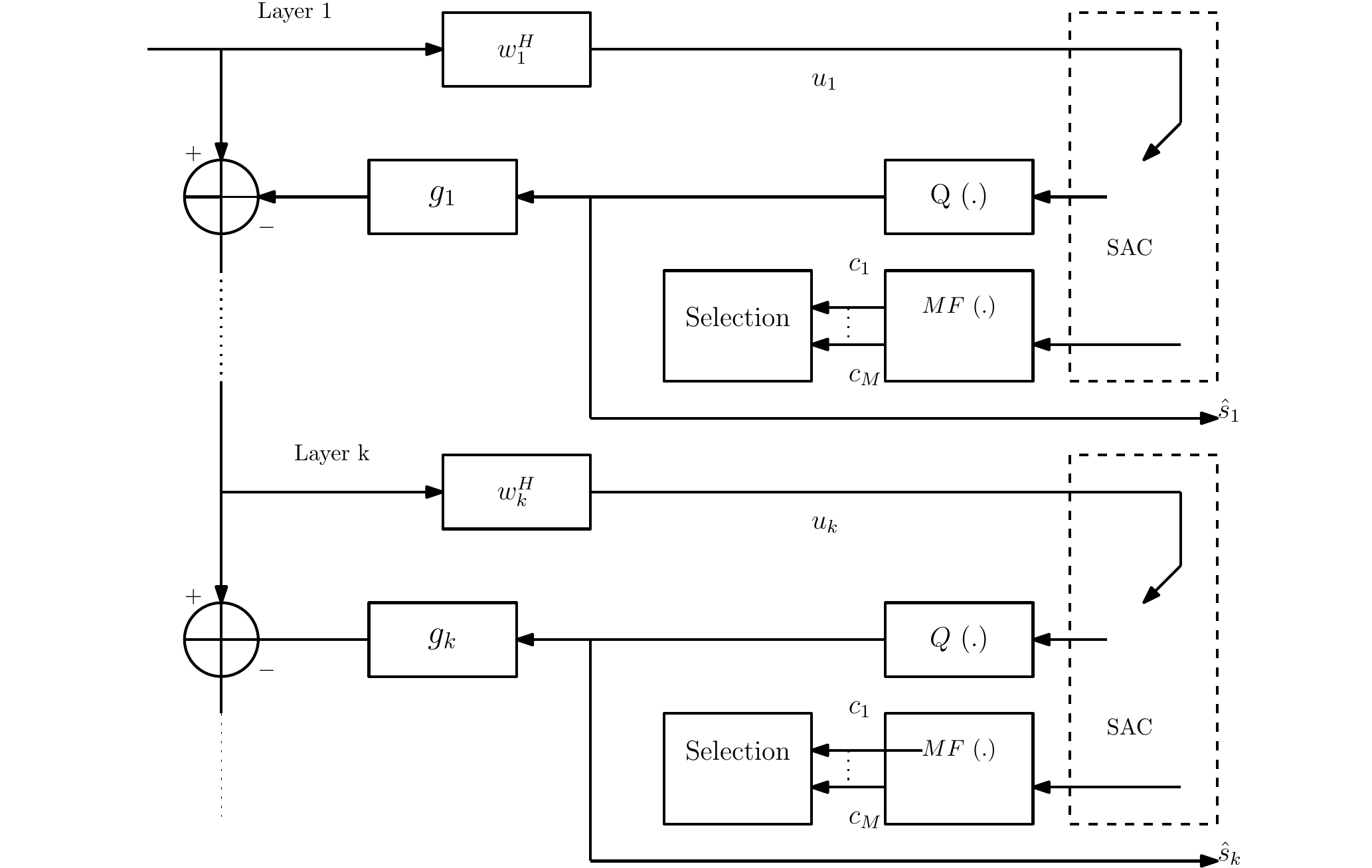}
\caption{Block diagram of a the Proposed MF-SIC detector.}\label{fig11}
\end{figure}

The block diagram of the proposed MF-SIC is shown in Fig. \ref{fig11}.  The design leverages on feedback diversity by choosing a set of constellation candidates when the previously detected symbol is considered to be unreliable \cite{r5}. A shadow area constraint (SAC) is introduced in order to obtain an optimal feedback candidate. This helps to  reduce the computation complexity in the search space, by avoiding it from growing exponentially. One of the positive attributes of such a selection criterion, is that there is no need for redundant processing when reliable decisions are made. Additionally, the proposed MF-SIC scheme mitigates error propagation that usually occurs when SIC-based approaches are used for detection.
The procedure for detecting $\hat{s}_{k}$ for the $k$-th user is described following a similar procedure presented in \cite{r9}. The $k$-th user soft estimate is obtained by $u_{k}=\mathbf{w}_{k}^{H}\check{\mathbf{y}}_{k}$ where the $L_{AP}\times1$ MMSE filter vector $\mathbf{w}_{k}=(\bar{\mathbf{G}}_{k}\bar{\mathbf{G}}^{H}_{k}+\frac{\sigma^{2}_{n}}{E_{s}}\mathbf{I})^{-1}\mathbf{g}_{k}$. $\bar{\mathbf{G}}_{k}$ represents the matrix obtained by  stacking the columns $k, k+1,...K$ of $\mathbf{G}$ and $\check{\mathbf{y}}_{k}=\mathbf{y}-\sum_{t=1}^{k-1}\mathbf{g}_{t}\hat{{s}}_{t}$ denotes the received vector after performing cancellation of the $k-1$ previously detected symbols. The soft estimate $u_{k}$ for each layer is examined by the SAC to determine if this decision is reliable according to
\begin{align}
    d_{k}=\vert u_{k}-\nu_{f}\vert,
\end{align}
where $\nu_{f}=\mathsf{arg}\min_{\mathbf{\nu_{f}}\in{\mathcal{A}}}
\biggl\{ \vert u_{k}-\nu_{f}\vert\biggr\}$ denotes the closest
constellation point to the $k$-th user soft estimate $u_{k}$. If
$d_{k}>d_{\text{th}}$ the decision is considered to be unreliable
and the selected constellation point is dropped into the shadow area
of the constellation map.   Parameter $d_{\text{th}}$ is the
predefined  threshold euclidean distance to guarantee reliability of
the selected symbol \cite{r5}. If the soft estimate $u_{k}$ is
deemed to be a reliable estimate for user $k$, the MF-SIC algorithm
performs a hard slice as in the conventional SIC approach
\cite{r5,r6}. In this case, $\hat{s}_{k}=Q(u_{k})$ is the estimated
symbol, where $Q(\cdot)$ is the quantization notation which maps to
the constellation symbol closest to  $u_{k}$.

Otherwise, the decision is deemed unreliable. In this case, a
candidate set $\mathcal{L}=\{c_{1}, c_{2},...,c_{m},...,c_{M}\}
\mathcal{A}$ is generated, which consists of the $M$  constellation
points closest to $u_{k}$. The number of candidate points $M$ is
given by the QPSK  symbols. As a result, there is a trade-off
between performance and complexity. The algorithm selects an optimal
candidate $c_{m,\text{opt}}$ from a pool of $\mathcal{L}$
candidates. As a result, the unreliable choice $Q(u_{k})$ is
substituted by a hard decision, and $\hat{s}_{k}=c_{m,\text{opt}}$
is obtained. It should be noted that the MF-SIC algorithm's
performance benefits are based on the assumption that
$c_{m,\text{opt}}$ is correctly selected. The following is a summary
of the MF-SIC selection algorithm: To begin, the selection vectors
${\phi}^{1},{\phi}^{2},...,{\phi}^{m},...{\phi}^{M}$ must be
defined.

The size of these selection vectors is equal to the number of the
constellation candidates that are used every time a decision is
considered unreliable. For example, for the $k$-th  layer,  a
$K\times 1$ vector ${\phi}^{m}  =\left [
\hat{s}_{1},...,\hat{s}_{k-1},
c_{m},\phi^{m}_{k+1},...,\phi^{m}_{q},...,
   \phi^{m}_{K} \right ]^T$  which is  a potential choice corresponding to $c_{m}$ in
the k-th user consists of the following items: (a) The previously
estimated symbols $\hat{s}_{1},\hat{s}_{2}, ..., \hat{s}_{k-1}$. (b)
The candidate symbol $c_{m}$ obtained from the constellation for
subtracting a decision that was considered unreliable $Q(u_{k})$ of
the k-th user. (c) Using (a) and (b) as the previous decisions,
detection of the next user data $k+1, ...,q,...,$ K-th is performed
by the  SIC approach. Mathematically, the choice $\phi^{m}$ is given
by \cite{r9}
\begin{align}
   \phi^{m}_{q}=Q(\mathbf{w}^{H}_{q}{\hat{{y}}}^{m}_{q}),
\end{align}
where the index $q$ denotes a given UE between the (k+1)-th and the K-th UE,
\begin{align}
    \hat{{y}}^{m}_{q}=\check{{y}}_{k}-\mathbf{g}_{k}c_{m}- \sum_{p=k+1}^{q-1}\mathbf{g}_{p}\phi^{m}_{p}.
\end{align}
A key attribute of the proposed MF-SIC algorithm is the same MMSE
filter  $\mathbf{w}_{k}$ that is used for all the constellation
candidates. Therefore, the proposed algorithm has the same
computational complexity as the conventional SIC. The optimal
candidate ${m,\text{opt}}$ is selected according to the local
maximum likelihood (ML) rule given by
\begin{align}
  {m,\text{opt}}=\mathsf{arg}\min_{1\leq m\leq M}\left \|{y}-\mathbf{G}{\phi}^{m}\right\|^{2}.
\end{align}

\section{Iterative detection and decoding}\label{IDD}

In this section, the MMSE- based  detectors are presented for the
IDD scheme as shown  in Fig. \ref{fig1}, consisting of a joint
detector and an LDPC decoder. Due to paper size limitation, the
operation is explained based on the MMSE detector given in
\eqref{equ10}. The received signal at the output of the filter,
contains the desired symbol, residual co-user interference and
noise. We use similar assumptions given  in
\cite{r1,r3,r10,dynovs,list_idd,msgamp,bimmsgamp} to approximate the
$\hat{s}_{k}$ as an AWGN channel given by
\begin{align}
  \hat{s}_{k}=\mu_{k}s_{k}+z_{k},
\end{align}
where $\mu_{k}=\mathbb{E}\{\hat{s}_{k}{s}^{*}_{k}\}$. The parameter $z_{k}$ is a zero-mean AWGN variable.
Using similar procedures as in \cite{r1}, the parameter $\mu_{k}$ is given by
\begin{align}
  \mu_{k}=\mathbf{g}^{H}_{k}\biggl(\frac{\sigma^{2}_{n}}{E_{s}}\mathbf{I}+\mathbf{G}\mathbf{\Delta}_{k}\mathbf{G}^{H}\biggr)^{-1}\mathbf{g}_{k}.
\end{align}
The variance of $\hat{s}_{k}$ variance $\lambda^{2}_{k}$ is given by
\begin{align}
  \lambda^{2}_{k} =\mathbb{E}\left \{\mu_{k}-\mu^{2}_{k}  \right \},
\end{align}
The extrinsic LLR computed by the detector for the $l$-th bit $l\in\left\{1,2,...,M_{c}\right\}$ of the symbol $s_{k}$ transmitted by the $k$-th user is \cite{r1,r3}
\begin{align}
  L_{D}\left ( b_{(k-1)M_{c}+l} \right )&=\log\frac{\sum _{s\in A^{+1}_{l}}f\left ( \hat{s}_{k}|s \right )P\left (s \right )}{\sum _{s\in A^{-1}_{l}}f\left ( \hat{s}_{k}|s \right )P\left (s \right )}\\&\notag-Lc\left ( b_{(k-1)M_{c}+l} \right ),
\end{align}
    where $A^{+1}_{l}$ is the set of $2^{Mc-1}$ hypothesis $s$ for which the $l$-th bit is $+1$. The a-priori probability $P(s)$ is given by \eqref{aprior_prob}. The approximation of the likelihood function  \cite{r3} $f(\hat{s}_{k}|s)$ is given by
    \begin{align}
        f\left ( \hat{s}_{k}|s \right )\simeq\frac{1}{\pi\lambda^{2}_{k}}\exp\left (-\frac{1}{\lambda^{2}_{k}} |\hat{s}_{k}-\mu_{k}s|^{2} \right ).
    \end{align}
The soft beliefs  are  exchanged between the proposed detectors and the decoder in an iterative manner.
The traditional  sum product algorithm (SPA) suffers from performance degradation caused by the tangent function especially in the error-rate floor region \cite{r10,dynovs}. Therefore, we use the box-plus SPA in this paper because it yields less complex approximations. The decoder is made up of two stages namely: The single parity check (SPC) stage and the repetition stage.
The LLR sent from check node $(CN)_{J}$ to variable node $(VN)_{i}$ is computed as
\begin{align}
    L_{j\longrightarrow i}=\boxplus{i^{'}\in N(j)\diagdown iL_{i^{'\longrightarrow j}}}.
\end{align}
As shorthand, we use  $L_{1}\boxplus L_{2}$ to denote the computation of $L(L_{1}\bigoplus L_{2})$. The LLR is computed  by
\begin{align}
L_{1}\boxplus L_{2}=&\log\left ( \frac{1+e^{L_{1}+L_{2}}}{e^{L_{1}}+e^{L_{2}}} \right ),\\\notag
  =&\mathrm{sign}(L_{1})\mathrm{sign}(L_{2})\min(\left | L_{1} \right |,\left | L_{2} \right |)\\\notag&+\log\left ( 1+e^{-\left |L_{1}+L_{2}  \right |} \right )-\log\left (1+e^{-\left |L_{1}-L_{2}  \right |}  \right ).
\end{align}
The LLR from $VN_{i}$ to $CN_{j}$ is given by
\begin{align}
    L_{i\longrightarrow j}=L_{i}+\sum_{j^{'}\in N(i)\backslash j}L_{j^{'}\longrightarrow i},
\end{align}
where the parameter $L_{i}$ denotes the LLR at $VN_{i}$, ${j^{'}\in N(i)\backslash j}$ denotes all CNs connected to $VN_{i}$ except $CN_{j}$. Alternative decoding strategies based on message passing are also possible \cite{tree,vfap,ids,kaids}.

\section{Simulation results and discussion}
\label{Num_Dis}

In this section, the BER performance of the  proposed soft detectors
is presented for the CF-mMIMO  and Col-mMIMO settings. The CF-mMIMO
channel exhibits high PL values due to  LS fading coefficients.
Thus, the SNR definition is given by
\begin{align}
    SNR=\frac{{\tr}(\sigma_{s}^{2}\mathbf{G}\mathbf{G}^{H})R}{L_{AP}K_{UE}\sigma_{w}^{2}},
\end{align}
\begin{figure*}[t]
\centering
\input{example_with_three_in_row.tex}
\caption{BER versus SNR for CF-mMIMO for (a) SIC, (b) MF-SIC and (c) PIC   with $L = 100$, $K = 40$, while varying the number of IDD iterations.}
\label{fig3}
\end{figure*}

The simulation parameters  are varied  as follows: We consider a
cell-free environment with  a square of dimensions $D\times D=$,
where $D=1$ km. Distances $d_{0}$ and $d_{1}$ are $10$ m and $50$ m,
respectively. $h_{AP}=15$m, $h_{u}=1.65$ m, $f=1900$ MHz,
$d_{th}=0.38$, LDPC code with code word length $256$ bits, $M=128$
parity check bits and $N-M$ message bits. The code rate
$R=\frac{1}{2}$. The maximum number of inner iterations is set to
$10$. We remark that different code designs are possible in this
context \cite{bfpeg,dopeg,memd,baplnc}.  The signal power
$\sigma^{2}_{s}=1$ and the simulations are run for $10^{3}$ channel
realizations. The modulation scheme used is quadrature phase shift
keying (QPSK).  Figure  \ref{fig3} presents the BER versus the SNR
as the number of IDD iterations are increased.  It can be visualized
that increasing IDD iterations yields lower BER. This is because
more  a posterior information is exchanged between the joint
detector and decoder as the iterations increase, which  improves the
system performance. The number of iterations do not cause any
marginal effect on the linear MMSE filter without cancellation
because there is no $\Delta_{k}$ in this filter which is needed for
the IDD to improve the performance. Fig. \ref{fig2} presents  the
BER versus the SNR for the CF-mMIMO system model for different
values of $L$ and the studied soft detectors for  two IDD
iterations.  The PIC and MF-PIC achieves the lowest BER values,
followed by MF-SIC, SIC, MMSE, in that order.  Additionally,
increasing $L$ and $K$ reduces the BER.  Also, the performance
benefit between conventional PIC and MF-PIC is negligible.

\begin{figure}
\centering
\input{BERAPUE}
\caption{BER versus SNR for CF-mMIMO  for the different detectors.}\label{fig2}
\end{figure}

%

\section{Concluding Remarks}\label{CO_FD}

In this paper, we have proposed list-based detectors for CF-mMIMO
architectures. Specifically, an IDD scheme using LDPC codes has been
studied. Additionally, the performance of the proposed MF-SIC/PIC
schemes has been compared with other detectors. The proposed MF-PIC
achieves lower BER values as compared to  SIC scheme.  Finally,
increasing IDD iterations significantly reduces the BER.

\end{document}

%% file: example_with_three_in_row.tex
\definecolor{mycolor1}{rgb}{0.00000,1.00000,1.00000}%
\definecolor{mycolor2}{rgb}{1.00000,0.00000,1.00000}%
\definecolor{mycolor3}{rgb}{0.83,0.69,0.22}%

\pgfplotsset{every axis label/.append style={font=\scriptsize
},
every tick label/.append style={font=\scriptsize
}
}

\begin{tikzpicture}[font=\scriptsize
] 
\begin{axis}[%
name=IF1,
width=0.45\columnwidth,
height=0.45\columnwidth,
scale only axis,
ymode=log,
xmin=-5,
xmax=15,
xlabel={SNR  [dB]},
xmajorgrids,
ymin=0.00001,
ymax=0.1,
ylabel={BER},
ymajorgrids,
legend entries={IDD $=1$,
IDD $=2$ 				
				},
legend style={fill=white, fill opacity=0.6, draw opacity=1,
text opacity =1,at={(0.03,0.03)}, anchor= south west,draw=black,fill=white,legend cell align=left,font=\scriptsize}
]

\pgfplotsset{
    every axis/.append style={
        extra description/.code={
            \node at (0.5,-0.35) {(a)};
        },
    },
}


\node [draw,fill=white,font=\tiny,anchor= north east] at (axis cs: 20, 0.5 ) { $M_{\text{Tx}} =1$, $M_{\text{Rx}}= 2$ };

\addplot+[smooth,color=black,solid,thick, every mark/.append style={solid} ,mark=square,
y filter/.code={\pgfmathparse{\pgfmathresult-0}\pgfmathresult}]
  table[row sep=crcr]{%
-5  0.030451953125	\\
0	0.01044775390625\\
5	0.00283125	\\
10	0.0005033203125	\\
15	0.00005986328125	\\
20	0.00000869140625	\\
25	0.00000205078125	\\
};

\addplot+[smooth,color=black,dashed, thick, every mark/.append style={dashed} ,mark=square,
y filter/.code={\pgfmathparse{\pgfmathresult-0}\pgfmathresult}]
  table[row sep=crcr]{%
-5	0.0256708984375	\\
0	0.00806884765625\\
5	0.0020681640625	\\
10	0.0004501953125	\\
15	0.00003984375	\\
20	0.00000224609375	\\
25	0.00000087890625	\\
};

\addplot[smooth,color=red,only marks, every mark/.append style={solid,fill=red!50}, mark=square]
  table[row sep=crcr]{%
	50 2\\
};\label{P0_Zz}

\addplot[smooth,color=black,only marks, every mark/.append style={solid,fill=black!50}, mark=square*]
  table[row sep=crcr]{%
	50 2\\
};\label{P1_Zz}

\addplot[smooth,color=red,only marks, every mark/.append style={solid, fill=red}, mark=star]
  table[row sep=crcr]{%
	50 2\\
};\label{P2_Zz}

\addplot[smooth,color=blue,only marks, every mark/.append style={solid, fill=blue!50}, mark=*]
  table[row sep=crcr]{%
	50 2\\
};\label{P3_Zz}

\addplot[smooth,color=mycolor3,only marks, every mark/.append style={solid, fill=mycolor3!50}, mark=triangle*]
  table[row sep=crcr]{%
	50 2\\
};\label{P4_Zz}

\node [draw,fill=white,font=\tiny,anchor= south west] at (axis cs: -10, 0.000004 ) {
\setlength{\tabcolsep}{0.5mm}
\renewcommand{\arraystretch}{.8}
};

\end{axis}

\begin{axis}[%
name=IF2,
    at={($(IF1.east)+(70,0em)$)},
		anchor= west,
width=0.45\columnwidth,
height=0.45\columnwidth,
scale only axis,
ymode=log,
xmin=-5,
xmax=15,
xlabel={SNR  [dB]},
xmajorgrids,
ymin=0.00001,
ymax=0.1,
ylabel={BER},
ymajorgrids,
legend entries={IDD $=1$,
IDD $=2$ 				
				},
legend style={fill=white, fill opacity=0.6, draw opacity=1,
text opacity =1,at={(0.03,0.03)}, anchor= south west,draw=black,fill=white,legend cell align=left,font=\scriptsize}
]

\pgfplotsset{
    every axis/.append style={
        extra description/.code={
            \node at (0.5,-0.35) {(b)};
        },
    },
}



\node [draw,fill=white,font=\tiny,anchor= north east] at (axis cs: 28, 0.5 ) { $M_{\text{Tx}} = M_{\text{Rx}}= 3$ };

\addplot+[smooth,color=red,solid, every mark/.append style={solid, fill=red}, mark=star,line width=1pt]
  table[row sep=crcr]{%
-5	0.0300701171875	\\
0	0.01023994140625\\
5	0.00277900390625	\\
10	0.0004951171875	\\
15	0.0000599609375	\\
20	0.00000859375	\\
25	0.00000205078125	\\
};

\addplot+[smooth,color=red,dashed, thick, every mark/.append style={dashed} ,mark=star,
y filter/.code={\pgfmathparse{\pgfmathresult-0}\pgfmathresult}]
  table[row sep=crcr]{%
-5	0.02534306640625	\\
0	0.00751435546875\\
5	0.001775	\\
10	0.00027626953125	\\
15	0.00002451171875	\\
20	0	\\
25	0.00000107421875	\\
};

\node[coordinate] (M2_edge) at  (axis cs:14,-10){};  

\end{axis}

\begin{axis}[%
name=IF3,
		at={($(IF2.east)+(70,0em)$)},
		anchor= west,
width=0.45\columnwidth,
height=0.45\columnwidth,
scale only axis,
ymode=log,
xmin=-5,
xmax=15,
xlabel={SNR  [dB]},
xmajorgrids,
ymin=0.00001,
ymax=0.1,
ylabel={BER},
ymajorgrids,
legend entries={IDD $=1$,
IDD $=2$ 				
				},
legend style={fill=white, fill opacity=0.6, draw opacity=1,
text opacity =1,at={(0.03,0.03)}, anchor= south west,draw=black,fill=white,legend cell align=left,font=\scriptsize}
]

\pgfplotsset{
    every axis/.append style={
        extra description/.code={
            \node at (0.5,-0.35) {(c)};
        },
    },
}



\node [draw,fill=white,font=\tiny,anchor= north east] at (axis cs: 28, 0.5 ) { $M_{\text{Tx}} =1$, $M_{\text{Rx}}= 3$  };

\addplot+[smooth,color=blue,solid, every mark/.append style={solid, fill=blue!50}, mark=*,line width=1pt]
   table[row sep=crcr]{%
-5	0.03062529296875	\\
0	0.0105828125\\
5	0.00287822265625	\\
10	0.00050859375	\\
15	0.000061328125	\\
20	0.00000908203125	\\
25	0.0000015625	\\
};

\addplot+[smooth,color=blue,dashed, thick, every mark/.append style={dashed} ,mark=o,
y filter/.code={\pgfmathparse{\pgfmathresult-0}\pgfmathresult}]
  table[row sep=crcr]{%
-5	0.01959814453125	\\
0	0.00513115234375\\
5	0.0009478515625\\
10	0.00011806640625	\\
15	0.00000869140625	\\
20	0	\\
25	0	\\
};

\end{axis}
\end{tikzpicture}%

%% file: BERAPUE.tex
%
%
%
\usetikzlibrary{positioning,calc}

\definecolor{mycolor1}{rgb}{0.00000,1.00000,1.00000}%
\definecolor{mycolor2}{rgb}{1.00000,0.00000,1.00000}%

\definecolor{mustard}{rgb}{0.92941,0.69020,0.12941}%

\definecolor{newpurple}{rgb}{0.5, 0 ,1}%

\definecolor{darkblue}{rgb}{0, 0.4470, 0.7410}

\pgfplotsset{every axis label/.append style={font=\footnotesize},
every tick label/.append style={font=\footnotesize},
every plot/.append style={ultra thick} 
}

\begin{tikzpicture}[font=\footnotesize] 

\begin{axis}[%
name=mse,
width  = 0.8\columnwidth,
height = 0.6\columnwidth,
scale only axis,
xmin  = -5,
xmax  = 15,
xlabel= {SNR [dB]},
xmajorgrids,
ymin=0.00001,
ymax=0.1,
ymode=log,
ylabel={BER},
ymajorgrids,
legend entries={MMSE,
					SIC,
                MF-SIC,
                PIC,
				MF-PIC,	
				},
legend style={fill=white, fill opacity=0.6, draw opacity=1,
text opacity =1,at={(0.05,0.05)}, anchor= south west,draw=black,fill=white,legend cell align=left,font=\scriptsize}
]

\addlegendimage{smooth,color=black,solid, thick, mark=x,
y filter/.code={\pgfmathparse{\pgfmathresult-0}\pgfmathresult}}
\addlegendimage{smooth,color=red,solid, thick, mark=square,
y filter/.code={\pgfmathparse{\pgfmathresult-0}\pgfmathresult}}
\addlegendimage{smooth,color=red,solid, thick, mark=star,
y filter/.code={\pgfmathparse{\pgfmathresult-0}\pgfmathresult}}
\addlegendimage{smooth,color=magenta,solid, thick, mark=o,
y filter/.code={\pgfmathparse{\pgfmathresult-0}\pgfmathresult}}
\addlegendimage{smooth,color=green,solid, thick, mark=diamond,
y filter/.code={\pgfmathparse{\pgfmathresult-0}\pgfmathresult}}




\addplot+[smooth,color=black,dashed,thick, every mark/.append style={solid} ,mark=x,
y filter/.code={\pgfmathparse{\pgfmathresult-0}\pgfmathresult}]
  table[row sep=crcr]{%
-5	0.044563720703125	\\
0	0.017219482421875\\
5	0.005014892578125	\\
10	0.0012421875	\\
15	0.000199951171875	\\
};

\addplot+[smooth,color=red,dashed, thick, every mark/.append style={solid} ,mark=square,
y filter/.code={\pgfmathparse{\pgfmathresult-0}\pgfmathresult}]
  table[row sep=crcr]{%
-5	0.041248047	\\
0	0.015239014\\
5	0.00435376	\\
10	0.001129639	\\
15	0.000248535	\\
20	3.13E-05	\\
25	1.00E-05	\\
};


\addplot+[smooth,color=red,dashed, thick, every mark/.append style={solid} ,mark=star,
y filter/.code={\pgfmathparse{\pgfmathresult-0}\pgfmathresult}]
  table[row sep=crcr]{%
-5	0.040718505859375	\\
0	0.014741943359375\\
5	0.0039609375	\\
10	0.000887451171875	\\
15	0.000165283203125	\\
};


\addplot+[smooth,color=magenta,dashed,thick, every mark/.append style={solid} ,mark=o,
y filter/.code={\pgfmathparse{\pgfmathresult-0}\pgfmathresult}]
   table[row sep=crcr]{%
-5	0.036134765625	\\
0	0.0122314453125\\
5	0.002912109375	\\
10	0.0005576171875	\\
15	0.00007470703125	\\
};


\addplot+[smooth,color=green,dashed, thick, every mark/.append style={solid} ,mark=diamond,
y filter/.code={\pgfmathparse{\pgfmathresult-0}\pgfmathresult}]
   table[row sep=crcr]{%
-5	0.036134765625	\\
0	0.0122314453125\\
5	0.002912109375	\\
10	0.0005576171875	\\
15	0.00007470703125	\\
};


\addplot+[smooth,color=black,solid, thick, every mark/.append style={solid} ,mark=x,
y filter/.code={\pgfmathparse{\pgfmathresult-0}\pgfmathresult}]
  table[row sep=crcr]{%
-5	0.029968	\\
0	0.011123\\
5	0.0027067\\
10	0.00062109\\
15	0.0001041\\
};
\addplot+[smooth,color=red,thick,solid, every mark/.append style={solid} ,mark=square,
y filter/.code={\pgfmathparse{\pgfmathresult-0}\pgfmathresult}]
  table[row sep=crcr]{%
-5	0.025308	\\
0	0.0084554\\
5	0.0019705	\\
10	0.00055996	\\
15	8.584e-05	\\
};

\addplot+[smooth,color=red,thick,solid, thick, every mark/.append style={solid} ,mark=star,
y filter/.code={\pgfmathparse{\pgfmathresult-0}\pgfmathresult}]
  table[row sep=crcr]{%
-5	0.024848	\\
0	0.0080958\\
5	0.0016336	\\
10	0.00036533	\\
15	7.3438e-05	\\
};
\addplot+[smooth,color=magenta,thick,solid, thick, every mark/.append style={solid} ,mark=o,
y filter/.code={\pgfmathparse{\pgfmathresult-0}\pgfmathresult}]
  table[row sep=crcr]{%
-5	0.019466	\\
0	0.0055334\\
5	0.00080293\\
10	0.00017549	\\
15	1.2891e-05	\\
};
\addplot+[smooth,color=green,thick,solid, thick, every mark/.append style={solid} ,mark=diamond,mark size=6,
y filter/.code={\pgfmathparse{\pgfmathresult-0}\pgfmathresult}]
  table[row sep=crcr]{%
-5	0.019466	\\
0	0.0055334\\
5	0.00080293\\
10	0.00017549	\\
15	1.2891e-05	\\
};
\addplot[smooth,color=black, thick, every mark/.append style={solid} ,mark=o,
y filter/.code={\pgfmathparse{\pgfmathresult-0}\pgfmathresult}]
  table[row sep=crcr]{%
	-1 -2\\
};\label{P33}

\addplot[smooth,color=black,thick,dashed, every mark/.append style={solid} ,mark=triangle, 
y filter/.code={\pgfmathparse{\pgfmathresult-0}\pgfmathresult}]
  table[row sep=crcr]{%
	-1 -2\\
};\label{P34}

\addplot[smooth,color=black,thick,dotted, every mark/.append style={solid} ,mark=square, 
y filter/.code={\pgfmathparse{\pgfmathresult-0}\pgfmathresult}]
  table[row sep=crcr]{%
	-1 -1\\
};\label{P35}

\node [draw,fill=white, fill opacity=0.6,draw opacity=1,
text opacity =1,at ={(-20,4.75)}, anchor= north west,draw=black,fill=white,font=\scriptsize]  {
\setlength{\tabcolsep}{0.5mm}
\renewcommand{\arraystretch}{.8}
\begin{tabular}{l}
\ref{P35}{\hspace{0.15cm} $N_r=10$}\\
\ref{P34}{\hspace{0.15cm} $N_r=7$}\\
\ref{P33}{\hspace{0.15cm} $N_r=4$}\\
\end{tabular}
};
\addplot[smooth,color=black,dashed,line width=1.0pt,mark=none,
y filter/.code={\pgfmathparse{\pgfmathresult-0}\pgfmathresult}]
  table[row sep=crcr]{%
	1 2\\
};\label{L1}

\addplot[smooth,color=black,solid,line width=1.0pt,mark=none,
y filter/.code={\pgfmathparse{\pgfmathresult-0}\pgfmathresult}]
  table[row sep=crcr]{%
	1 2\\
};\label{L2}

\node [draw,fill=white,font=\footnotesize,anchor= north  east] at (axis cs: 14,0.08) {
\setlength{\tabcolsep}{0.5mm}
\renewcommand{\arraystretch}{1}
\begin{tabular}{l}
\ref{L1}{~$K=16, L=64$}\\
\ref{L2}{~$K=40, L=100$}\\
\end{tabular}
};

\end{axis}
\end{tikzpicture}